\documentclass[twocolumn]{aastex63}

\received{ }
\revised{ }
\accepted{ }
\submitjournal{ApJ}

\shorttitle{Failed eruption of a helical prominence}
\shortauthors{Xu et al.}

\usepackage{graphicx}

\begin{document}

\title{Multi--wavelength Observation of a Failed Eruption from a Helical Kink-unstable Prominence}
\correspondingauthor{Haiqing Xu}
\email{xhq@bao.ac.cn}
\author{Haiqing Xu}
\affil{Key Laboratory of Solar Activity, National Astronomical
Observatories, Chinese Academy of Sciences, Beijing 100101, China}
\author{Jiangtao Su}
\affil{Key Laboratory of Solar Activity, National Astronomical
Observatories, Chinese Academy of Sciences, Beijing 100101, China}
\affil{School of Astronomy and Space Science, University of Chinese Academy of Sciences, Beijing, China}
\author{Jie Chen}
\affil{Key Laboratory of Solar Activity, National Astronomical
Observatories, Chinese Academy of Sciences, Beijing 100101, China}
\author{Guiping Ruan}
\affil{Institute of Space Sciences and School of Space Science
and Physics, Shandong University, Weihai 264209, China}
\author{Arun Kumar Awasthi}
\affil{CAS Key Laboratory of Geospace Environment, Department of Geophysics and
Planetary Sciences, University of Science and Technology of China, Hefei 230026,
China}
\author{Hongqi Zhang}
\affil{Key Laboratory of Solar Activity, National Astronomical
Observatories, Chinese Academy of Sciences, Beijing 100101, China}
\author{Mei Zhang}
\affil{Key Laboratory of Solar Activity, National Astronomical
Observatories, Chinese Academy of Sciences, Beijing 100101, China}
\affil{School of Astronomy and Space Science, University of Chinese Academy of Sciences, Beijing, China}
\author{Kaifan Ji}
\affil{Yunnan Observatories, Chinese Academy of Sciences, Kunming, 650216, China}
\author{Yuzong Zhang}
\affil{Key Laboratory of Solar Activity, National Astronomical
Observatories, Chinese Academy of Sciences, Beijing 100101, China}
\author{Jiajia Liu}
\affil{Astrophysics Research Centre, School of Mathematics $\&$ Physics, Queen's University Belfast, Belfast, BT7 1NN, UK}

\begin{abstract}
Multi-wavelength observations of prominence eruptions provide an opportunity to uncover the physical mechanism of the triggering and the evolution process of the eruption. In this paper, we investigated an erupting prominence on October 14, 2012, recorded in H$\alpha$, EUV, and X-ray wavelengths. The process of the eruption gives evidences on the existence of a helical magnetic structure showing the twist converting to writhe. \textbf{The estimated twist is} $\sim6\pi$ (3 turns), exceeding the threshold of the kink instability. The rising plasma reached a high speed at 228 km s$^{-1}$, followed by a sudden rapid acceleration at 2715 m s$^{-2}$, and synchronous with a solar flare. Co-spatial cusp shaped structures were observed in both AIA 131 \AA\ and 94 \AA\ images, signifying the location of the magnetic reconnection. The erupted flux rope finally undergone a deceleration with a maximum value of 391 m s$^{-2}$, which is larger than the free-fall acceleration on the Sun (273 m s$^{-2}$), suggesting that the eruption finally failed, possibly due to an inward magnetic tension force.
\end{abstract}


\section{Introduction}

Prominence eruptions are large-scale eruptive phenomena that are frequently observed in the corona. Observations show that ejective prominences (filaments) are often associated with coronal mass ejections (CMEs) and flares \citep[e.g.,][]{Web76, Zho06, Fil08, Lie13}. This makes the investigations of these events crucial for understanding sun-earth connection. There are also prominence eruptions that are not accompanied by CMEs. In these events, the associated coronal structure remains in the corona, with the prominence material often falling back to the chromosphere. These events are called failed (confined) eruptions.

While successful eruptions are important as their association with CMEs, failed eruptions can also shed light on the pre- and post-eruption magnetic field configurations. \citet{Ji03} analyzed a typical failed filament eruption and found that the energy release and reconnection point may occur at the location above the filament during its acceleration phase. Recently, using high spatial and temporal resolution data from ground-based and space-borne facilities, several observational studies of confined eruptions have been done \citep[e.g.,][]{She12, Net12, Liu14, Yang14, Li15, Xue16a}. Although the pre-eruptive structure is usually very complicate, the helical structure often becomes prominent during the acceleration phase \citep{vrs91}. \citet{De99} observed internal helical structures of CMEs, and interpreted them to be magnetic flux ropes. \citet{Che11} presented an unambiguous observation of a flux rope in the formation phase in the low corona. The magnetic flux rope is a key feature in the prominence \citep[e.g.,][]{Ali07, Guo10, Kum10, Kol12, Che14, Yang15}. This observational \textbf{phenomenon} has also motivated many scientists to include magnetic flux ropes in their numerical simulations of solar eruptions  \citep[e.g.,][]{Lin98, Tit99, Ama00}.

The mechanism of both ejective and failed eruptions is not yet fully understood, owing to the fact that the underlying magnetic structure is poorly known. It has been suggested that magnetic reconnection can trigger filament eruptions and CMEs \citep{Fey95, Wan99}. However, it is difficult to observe magnetic reconnection directly because of the very small size of the diffusion region. Mostly it is observed through the consequences of the reconnection, such as the the heating of the corona \citep[e.g.,][]{Zha15}, inflows \citep{Yok01, Su13} and/or outflows indicated from both imaging and spectral observations \citep[e.g.,][]{Sav10, Tia14, Ree15, Hon16}. Recently, \citet{Li16} gave a clear observational evidence of magnetic reconnection between an erupting filament and its nearby coronal loops. \citet{Xue16b} presented comprehensive observational evidence of reconnection between a set of chromospheric fibrils and the threads of an erupting filament, which leads to the eventual untwisting of a flux rope.

Another widely accepted mechanism of solar eruption is the kink instability. It was initially suggested as the trigger of both confined and ejective prominence eruptions by \citet{Sak76}. Magneto-hydrodynamics (MHD) simulation \citep{Tor05} confirmed that the helical kink instability of a twisted magnetic flux rope can be the mechanism of the initiation and the initial driver of solar eruptions, and the decrease of the overlying field with height is a main factor in deciding whether the instability leads to a confined event or a CME. Several observational studies also show that the kink instability is the driver of filament eruptions \citep[e.g.,][]{Wil05, Rus05}. Of course, there are also some other mechanisms for solar eruptions, such as sunspot rotation and shearing motions \citep[e.g.,][]{Tia06, Ruan14,Chen15}, wave disturbance \citep{Uch74}, critical twist configuration \citep{Vrs88} etc.

The study of twist provides an effective tool for analyzing the structure and stability of prominences. The twist of a flux rope is determined by the number of the turns of the magnetic field line when counting from one footpoint to the other. The critical twist number for kink instability varies at different conditions. \citet{Hoo79} found that, for a force-free uniformly-twisted flux rope, there is a critical twist of 3.3$\pi$,  exceeding which the field will become kink-unstable.  \citet{Tor03} simulated the formation of a twisted magnetic flux rope and found that the critical twist number is between $2.5\pi$ and $2.75\pi$, for a set of different parameters. Other simulations \citep{Fan05, Tor05} also show that a magnetic flux rope becomes kink-unstable if the twist exceeds a critical value of $2\pi$. Models of flux ropes in an external field give values closer to $3.5\pi$ for the critical twist \citep{Fan03, Fan04, Tor04}. This value even increases with increasing aspect ratio of the loops involved \citep{Bat01, Tor04}. So it should be interesting and important to check the number of the critical twist from observations.

In this paper, we analyze a prominence eruption using observations from Solar Magnetism and Activity Telescope (SMAT) at Huairou Solar Observing Station (HSOS) of the National Astronomical Observatories of China and Atmospheric Imaging Assembly (AIA) onboard the Solar Dynamics Observatory (SDO). We also use the Solar Terrestrial Relations Observatory (STEREO) mission observations to combine limb and on-disk observations. We aim at investigating the morphology as well as the kinematic and helicity evolutions of the prominence during its eruption. In Section 2, we describe the observations. In Section 3, we present the main results. Discussions and conclusions are outlined in Section 4.

\section{Observations and data analysis}

\begin{figure*}
\includegraphics[width=1.0\textwidth]{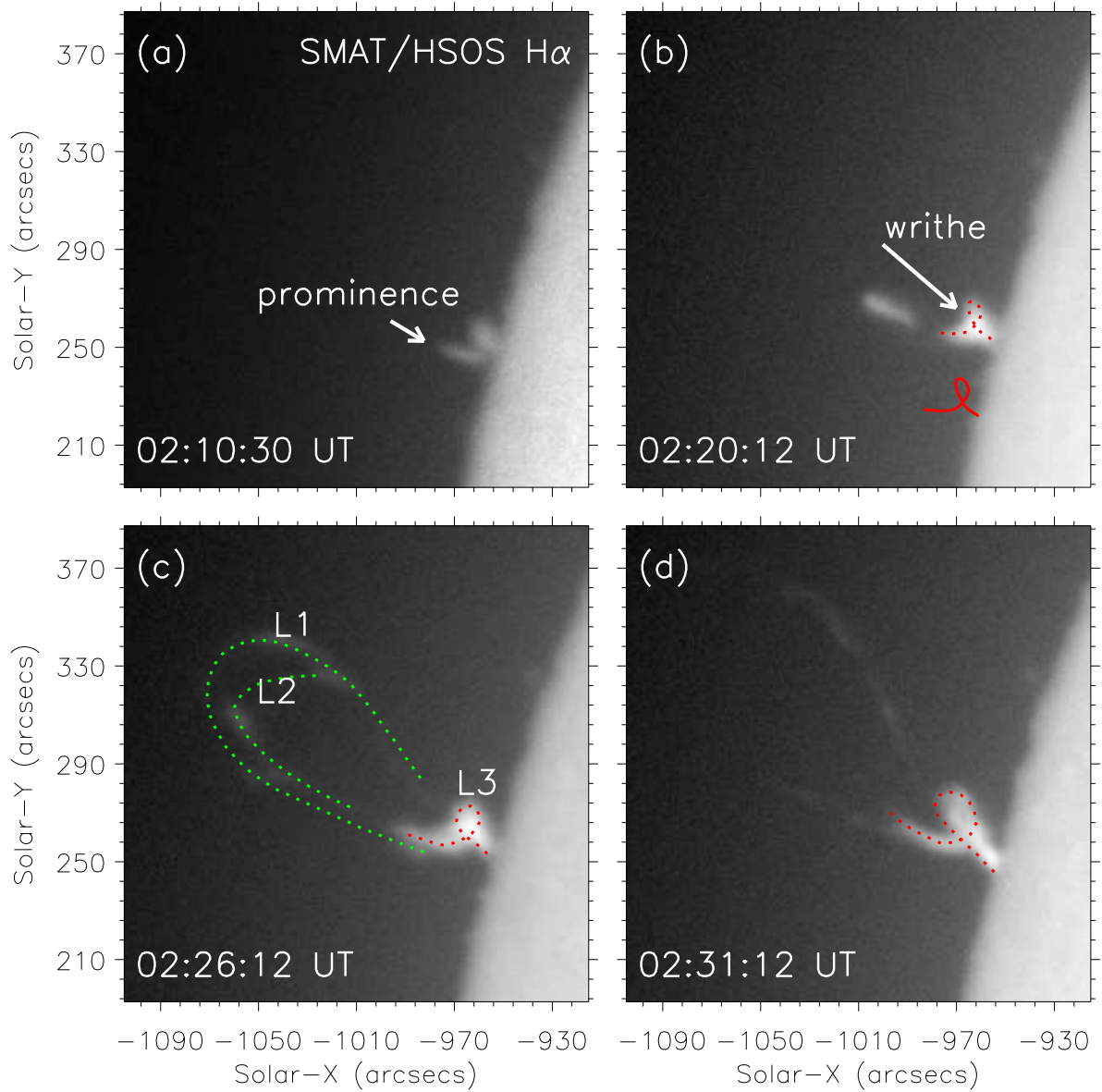}
\caption{Evolution of the prominence in H$\alpha$ images. The red and green dotted lines show the profile of prominence. The diagram of writhe in panel (b) is shown by the red solid line. See also supplementary Movie 1. }\label{fig:xu1}
\end{figure*}

\begin{figure*}
\includegraphics[width=1.0\textwidth]{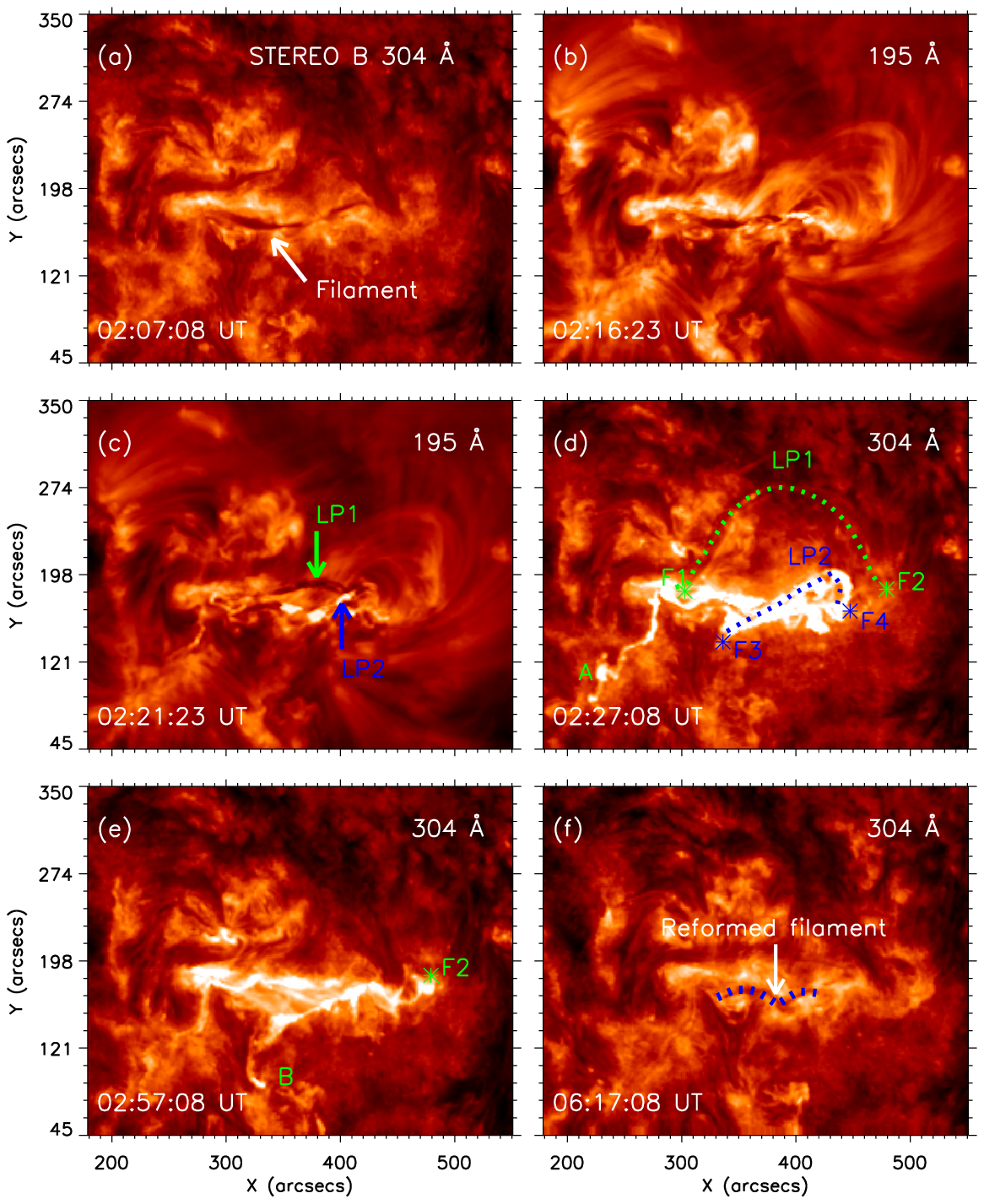}
\caption{Evolution of the filament in STEREO EUVI B 304 {\AA} and 195{\AA} (Panels (b) and (c)) images. White arrow in Panel (a) points to the filament before eruption and in Panel (f) the reformed filament. The green and blue arrows in Panel (c) point to LP1 and LP2 respectively.  The green and blue dotted lines in Panel (d) show the loops LP1 and LP2 respectively. The green (F1 and F2) and blue (F3 and F4) star symbols show the two footpoints of LP1 and LP2 respectively.  `A' in Panel (d) and `B' in Panel (e) denote the locations of the bright ribbons. See also supplementary Movie 2. }\label{fig:xu3}
\end{figure*}

A prominence was observed on the north-east solar limb at 02:00 UT on 14 October 2012. It remained stable until 02:10 UT and then started to rise slowly, as shown in H$\alpha$ image observed by SMAT (Figure \ref{fig:xu1}). The full solar disk H$\alpha$ (6562.81 \AA) telescope of SMAT is operated by a collimated optics of 20 cm aperture and 180 cm effective focal length. The image size of the telescope is 9 mm $\times$ 9 mm, and the size of the CCD is 2029$\times$2044 pixels. The spatial resolution is better than $2^{\prime\prime}$ \citep{Zha07} and cadence of the acquired images is 1 second.

The SDO/AIA \citep{Lem12} provides multiple, simultaneous high-resolution full-disk images of the transition region and the corona. The AIA observes seven EUV, two UV, and one visible-light wavelength bands. The spatial and temporal resolutions of AIA are $1.2^{\prime\prime}$ (pixel size is $0.6^{\prime\prime}$) and 12 s, respectively. The field of view is 1.3 $R_{\odot}$. The temperature response of the EUV emission line covers a wide range of material heated, from 1 MK to 20 MK.

The event was also observed as a filament eruption in the Extreme Ultraviolet Imager (EUVI) onboard STEREO B. The two STEREO \textbf{spacecrafts}, launched in 2006 October, were placed into orbits around the Sun similar to Earth's, but with STEREO-A ahead of Earth in its orbit and STEREO-B behind, and with both spacecrafts gradually drifting away from Earth at a rate of about $22^{\circ}$ per year \citep{How08}. On 14 October 2012, the separation angle between STEREO A (B) and Earth was $126.3^{\circ} (119.8^{\circ})$, and STEREO A and B was $113.9^{\circ}$. The EUVI detector has 2048 x 2048 pixels and a pixel size is $1.6^{\prime\prime}$, and observes in four spectral channels that span the 0.1 to 20 MK temperature range \citep{Wue04}. We used the 304 \AA\ images with a cadence of 10 minutes to study the evolution of the filament.

For H$\alpha$ data, all images are aligned to the image observed at 02:00 UT by computing the cross correlation using the properties of the Fourier transform. We downloaded the level 1 AIA data, then derotated and co-aligned the AIA images to the time of 02:10 UT.

Ramaty High Energy Solar Spectroscopic Imager (RHESSI) \citep{Lin02} provides high resolution imaging and spectroscopy of X-ray (6 keV) to gamma-rays (17 MeV), to diagnose the heated plasma and accelerated electrons. For this event,  the RHESSI observations were available only from 02:50 UT. Here we used the hard X-ray emission primarily for locating the reconnection site and the source of thermal and non-thermal emission.

\begin{figure*}
\includegraphics[width=1.0\textwidth]{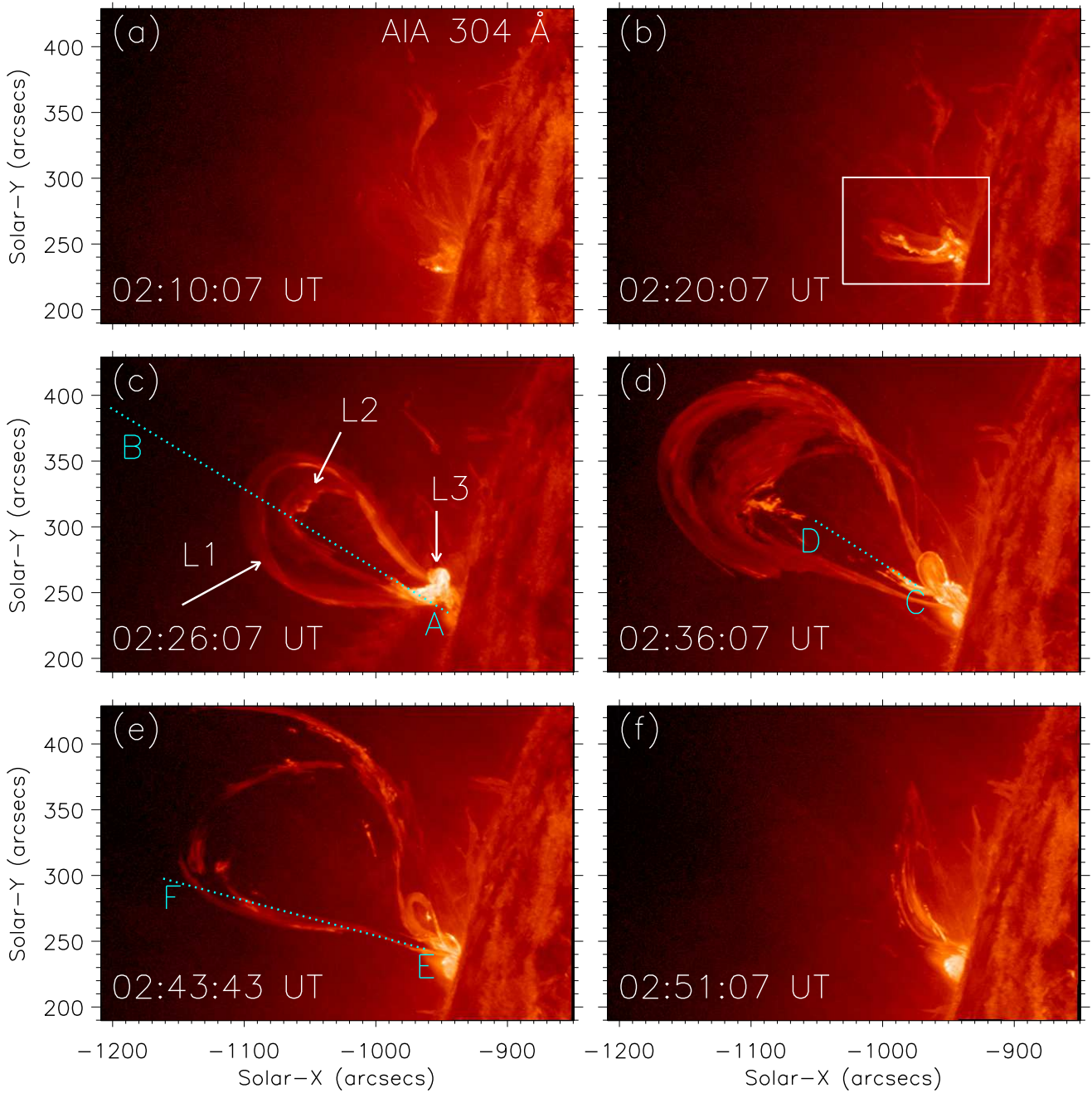}
\caption{The evolution of prominence in AIA 304 \AA\ images. White box in panel (b) shows the region to be used in Figure \ref{fig:xu4}. Three slices, used for synthesizing time-distance map in Figure \ref{fig:xuth}, are indicated by cyan dotted lines. See also supplementary Movie 3.
}\label{fig:xu2}
\end{figure*}

\begin{figure*}
\includegraphics[width=0.4\textwidth,angle=90]{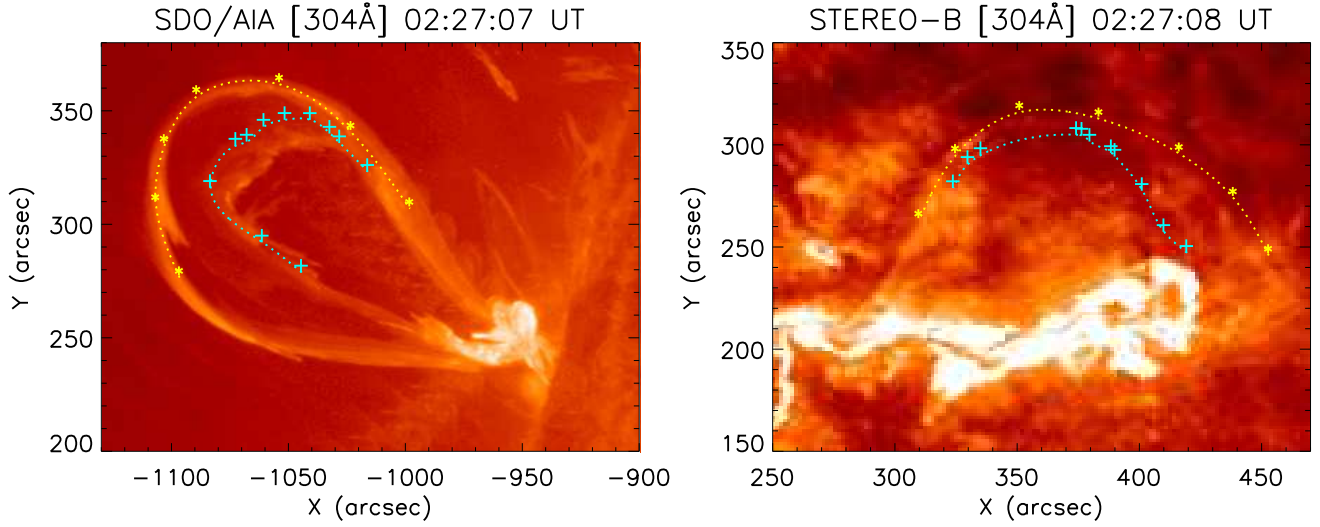}
\caption{\textbf{SDO/AIA 304 \AA\ and STEREO/EUVI 304 \AA\ images showing results from 3D triangulation exercise using SCC\_MEASURE.PRO technique. Symbols (`*' on L1 and `+' on L2) represent the identified points on respective images, while the dotted lines are spline-fit to the identified coordinates.}
}\label{fig:3d}
\end{figure*}

\begin{figure*}
\includegraphics[width=1.0\textwidth]{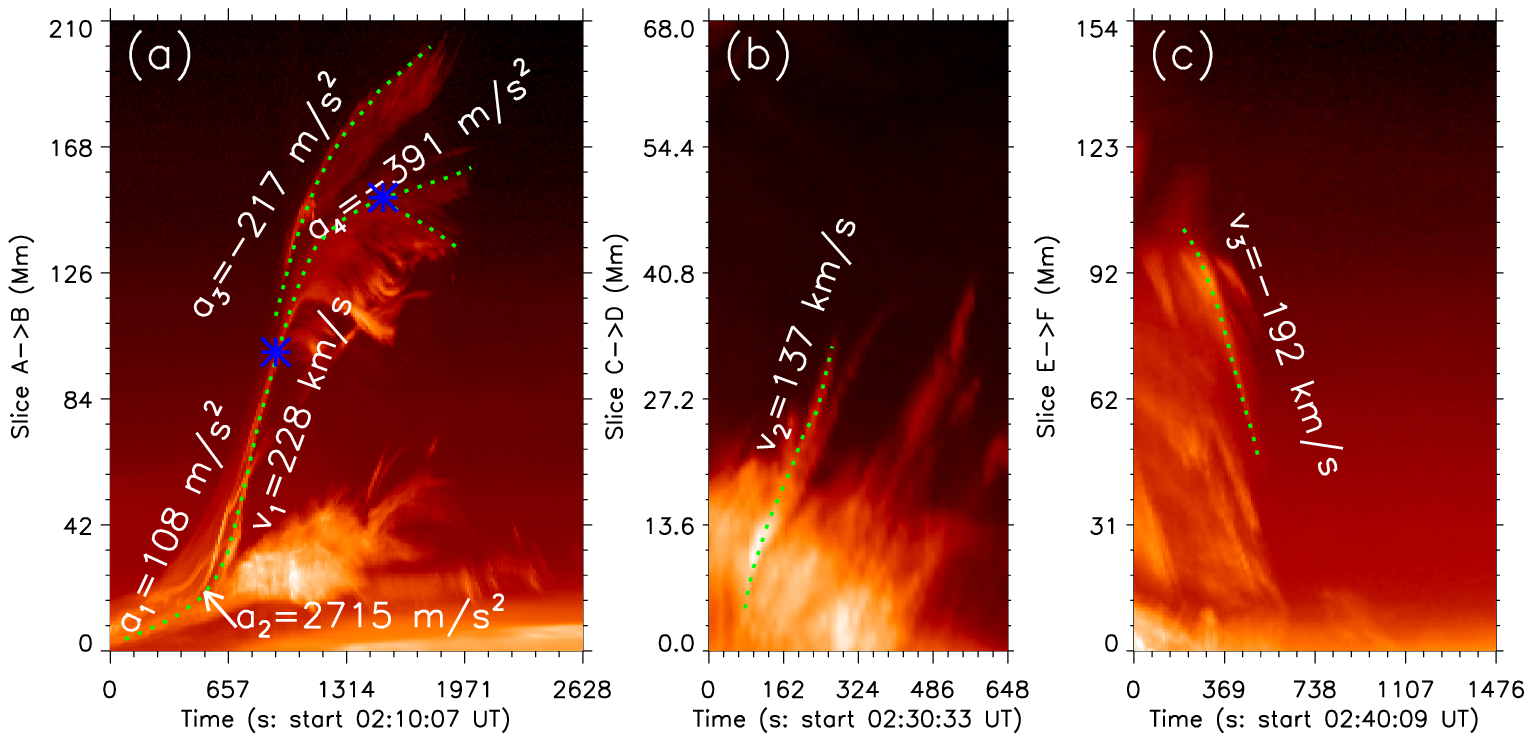}
\caption{Time-distance plots show flux evolution along three slices: (a) A$\rightarrow$B, (b) C$\rightarrow$D, (c) E$\rightarrow$F. Green dotted lines are used to determine the velocity and acceleration/deceleration of flux rope. Values of velocity and acceleration/deceleration at each phase are marked on the panels.
}\label{fig:xuth}
\end{figure*}

\section{Results}

\subsection{The morphology and kinematic evolution}

The prominence appeared as two bright structures in H$\alpha$ observations at 02:00:30 UT, wherein the lower part first started to rise up at 02:10:30 UT, as shown by the arrow in Figure \ref{fig:xu1}(a). At 02:20:12 UT (Figure \ref{fig:xu1}(b)), it became elongated and writhed as indicated by the red dotted line (a diagram for this structure is shown by the red solid line) while the clockwise rotation of the prominence lower part can be seen in the animation (see supplementary Movie 1). At 02:26:12 UT, the prominence \textbf{comprised of three parts}: the large faint \textbf{loops marked by L1 and L2} (green dotted line in Figure \ref{fig:xu1}(c)) and bright writhed core \textbf{marked by L3} (red dotted line in Figure \ref{fig:xu1}(c)). These \textbf{three parts} appeared to be closely connected in the middle-lower part of the prominence.  The writhe is more obvious at this time. On the other hand, the upper footpoint of the loop is too faint to be traced. With the loop rising and expanding, at 02:31:12 UT, the top of the eruptive loop was out of the field of view of H$\alpha$ images (Figure \ref{fig:xu1}(d)), while writhed part uplifted and became clearer.  Subsequently, the prominence material fell down at 02:38:12 UT. At about 02:59:11 UT, the eruption process ended.

In H$\alpha$ images, the footpoints of the prominence were not very clear. Therefore we attempted to derive such information from the EUVI/STEREO observation. The prominence was seen as a filament in EUVI B 304 \AA\ and 195 \AA\  images. It \textbf{appeared} to be in the quiet phase at 02:07:08 UT (Figure \ref{fig:xu3}(a)). From the sequence of 195 \AA\ images (available with 5-minute time cadence, see Figure \ref{fig:xu3}(b) and (c)), filament splitting was discernible since 02:21:23 UT. At 02:27:08 UT, the two branches of erupting filament completely separated: large faint arcade LP1 (green dotted line in Figure \ref{fig:xu3}(d)) located above the small helical bright core LP2 (blue dotted line in Figure \ref{fig:xu3}(d)). The eruption appeared to start with the writhe of the bright core (LP2), which appeared to drive the eruption of LP1. LP1 rose to a much higher height than LP2. The two legs of LP1 anchored at footpoints F1 and F2. Non-uniform thread density can be seen in at least three locations along LP1,  which indicates the twisted nature of loop LP1 (Figure 2(c)). The two legs of LP2 anchored at footpoints F3 and F4. The material moved to region `A'  due to the flaring activity (see supplementary Movie 2). As the plasma fell back, the span of brightened region increased around F2 and `B' due to energy deposited by the draining plasma, and the flare ribbons separated gradually (Figure \ref{fig:xu3}(e)). The filament reformed almost at the same location (Figure \ref{fig:xu3} (f)) which indicated that the filament eruption failed.

\begin{figure*}
\includegraphics[width=1.0\textwidth]{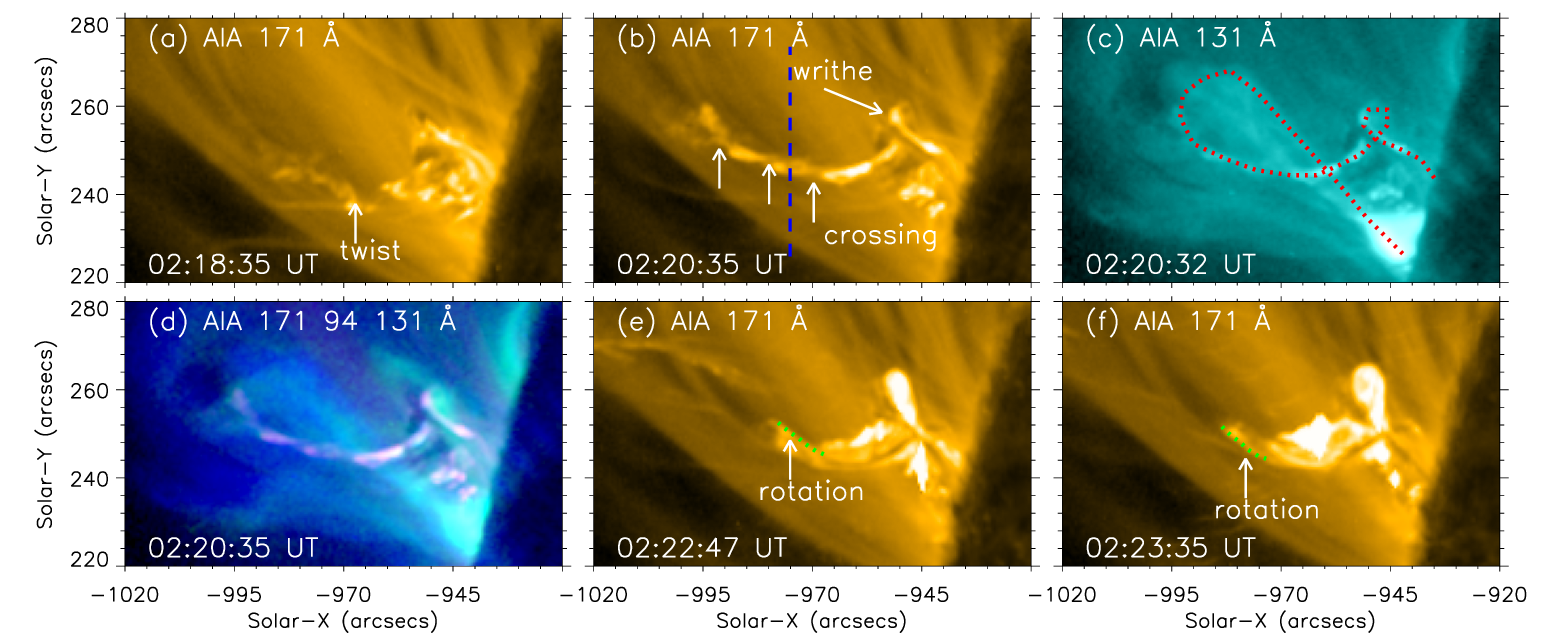}
\caption{Evolution of the twist in AIA images. Panels (a), (b), (e) and (f) are 171 \AA\ images. Panel (c) is 131 \AA\ image. Panel (d) is a composite image of the AIA 171 \AA\ (red), 94 \AA\ (green) and 131 \AA\ (blue) passbands. Their FOV is marked by the white box in Figure \ref{fig:xu2} (b). White arrows in panels show: (a) the twist, (b) writhe and threads crossing,  (c) kink structure (red dotted line), respectively. Green dotted lines in Panels (e) and (f) mark the rotation of the threads. See also supplementary Movie 4.}\label{fig:xu4}
\end{figure*}

\begin{figure*}
\centerline{\includegraphics[width=1.0\textwidth]{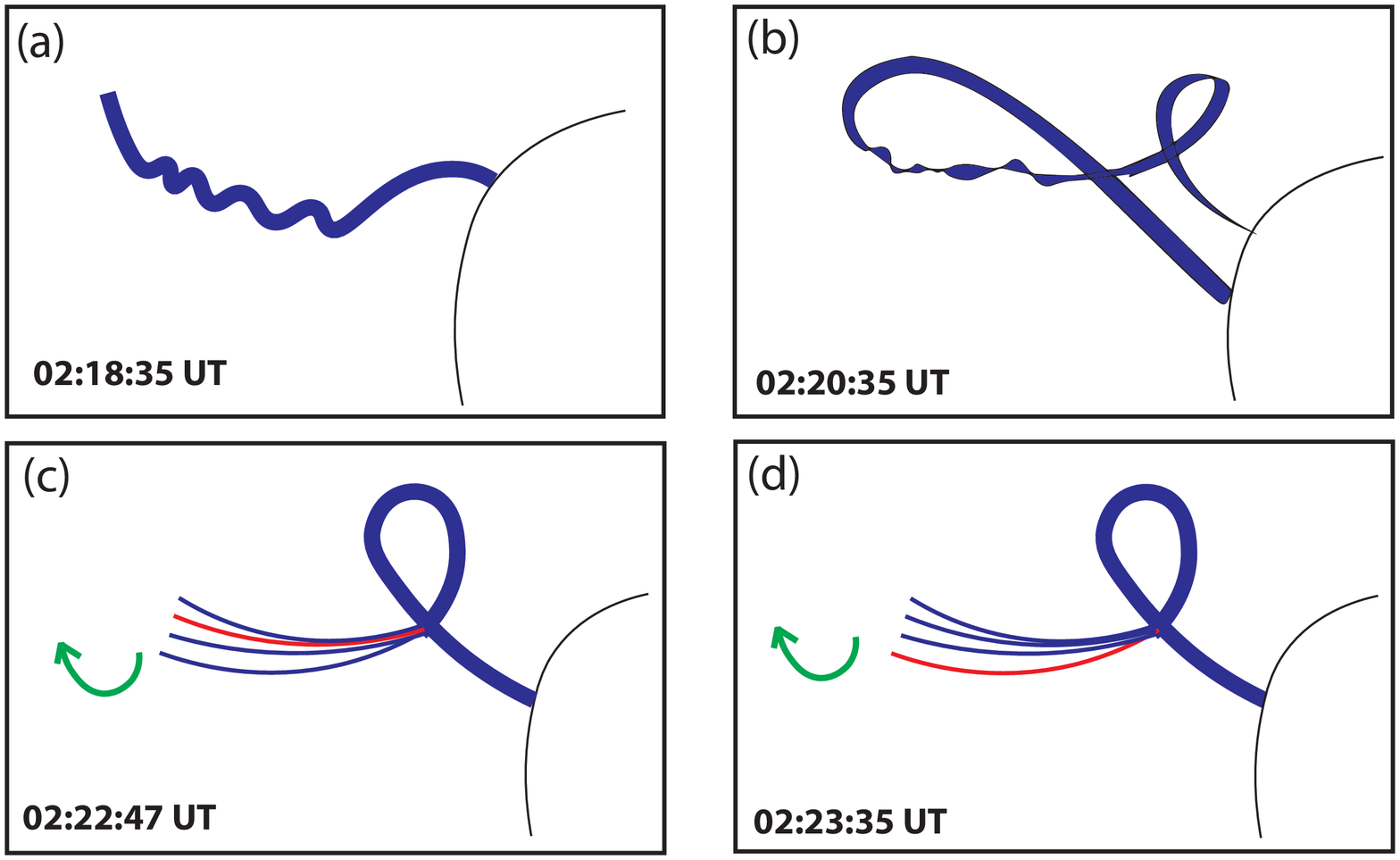}}
\caption{The sketch of twist evolution and rotation. (a): Corresponding to Figure \ref{fig:xu4}(a). (b) Corresponding to Figure \ref{fig:xu4}(d). (c) and (d): Corresponding to Figure \ref{fig:xu4}(e) and (f). The thin lines represent part of the prominence's treads. The green arrows show the rotation direction traced by the red thin line. The black curves in each panel represent the solar limb and thick blue curves represent the prominence.} \label{fig:xu11}
\end{figure*}

The prominence appeared in all seven AIA EUV channels at 02:10 UT, in agreement with the H$\alpha$ observations. We used AIA 304 \AA\ images to quantify the evolution of the prominence in EUV wavelength. There is an obvious twist and writhe at the core of the rising loop at 02:20:07 UT (Figure \ref{fig:xu2}(b)). This helical structure can also be seen in AIA hot \textbf{passbands}, which may relate to a magnetic flux-rope. The helical structure kept rising and expanding. \textbf{From 02:26:07 UT to 02:36:07 UT (Figure \ref{fig:xu2}(c) and (d))}, the prominence separated into \textbf{three parts: two large diffused loops marked by L1 and L2} with twisted fine threads \textbf{and a writhed structure marked by L3}. The south leg almost bifurcated into two loops \textbf{(L1 and L2)} with its footpoint attached to the surface. But the north leg only separated in the upper part and its footpoint could not be seen, suggesting that it was in the back side.  In the lower part of the south leg, strong brightening appeared which is indicative of the energy release following the magnetic reconnection.  The loop rose, expanded and unwrapped in due course of its evolution. the prominence became more diffused and some threads of L1 broke as shown in Figure \ref{fig:xu2}(d). The fast upward flow brought the hot plasma into the higher corona. This flow was more evident in hot EUV channels. After 02:43:43 UT, the loops disrupted and the mass started to fall down along the two legs of the prominence (Figure \ref{fig:xu2}(e)). The draining plasma deposited its hot materials to the solar surface and hence appeared brightened (Figure \ref{fig:xu2}(f)). From the animation (see supplementary Movie 3), clockwise rotation of L2's south-lower part can be seen during the rising phase \textbf{viewed from the top of L2.}

\textbf{In order to distinguish different loop structures, we retrieve the three-dimensional (3D) location of the filament derived by a triangulation technique called tie point \citep{Inh06}. We use the routine SCC\_MEASURE.PRO in the SSW package, which returns the location and height of the filament in 3D. The result is shown in Figure \ref{fig:3d}. Symbols (`*' on L1 and `+' on L2) represent the identified points on respective images, while the dotted lines are spline-fit to the identified coordinates. It confirms that the L1 and L2 marked in Figure \ref{fig:xu1}(c) and Figure \ref{fig:xu2}(c) are corresponding to large fainter filament arch (LP1 and it's vicinity) in Figure \ref{fig:xu3}(d), and L3 is corresponding to LP2 in Figure \ref{fig:xu3}(d).}

All of the aforementioned observations suggest that this region might be the so-called double-decker filament system as reported by \citet{Liu12} and \citet{Kli14}. The lower filament became unstable first due to kink instability, which then triggered the eruption of the upper filament.

\begin{figure*}
\includegraphics[width=1.0\textwidth]{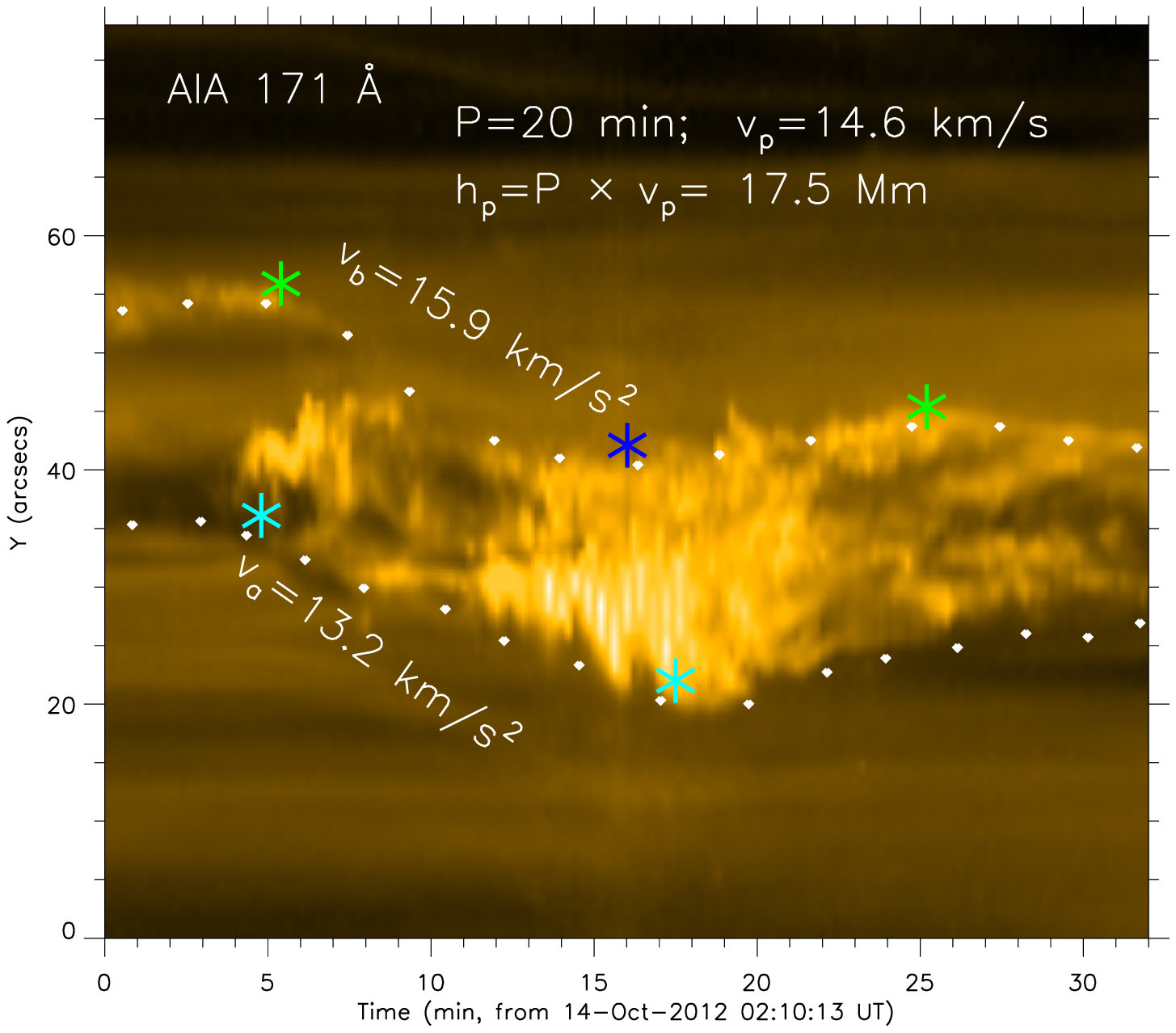}
\caption{Time-distance plot along the blue dashed line in Figure \ref{fig:xu4}(b), indicating the rotation of the prominence. The estimated values of period $P$, average speed $v_{p}$ and pitch of screw thread $h_{p}$ are listed on the panel.}\label{fig:xu12}
\end{figure*}

\begin{figure*}
{\includegraphics[width=1.0\textwidth]{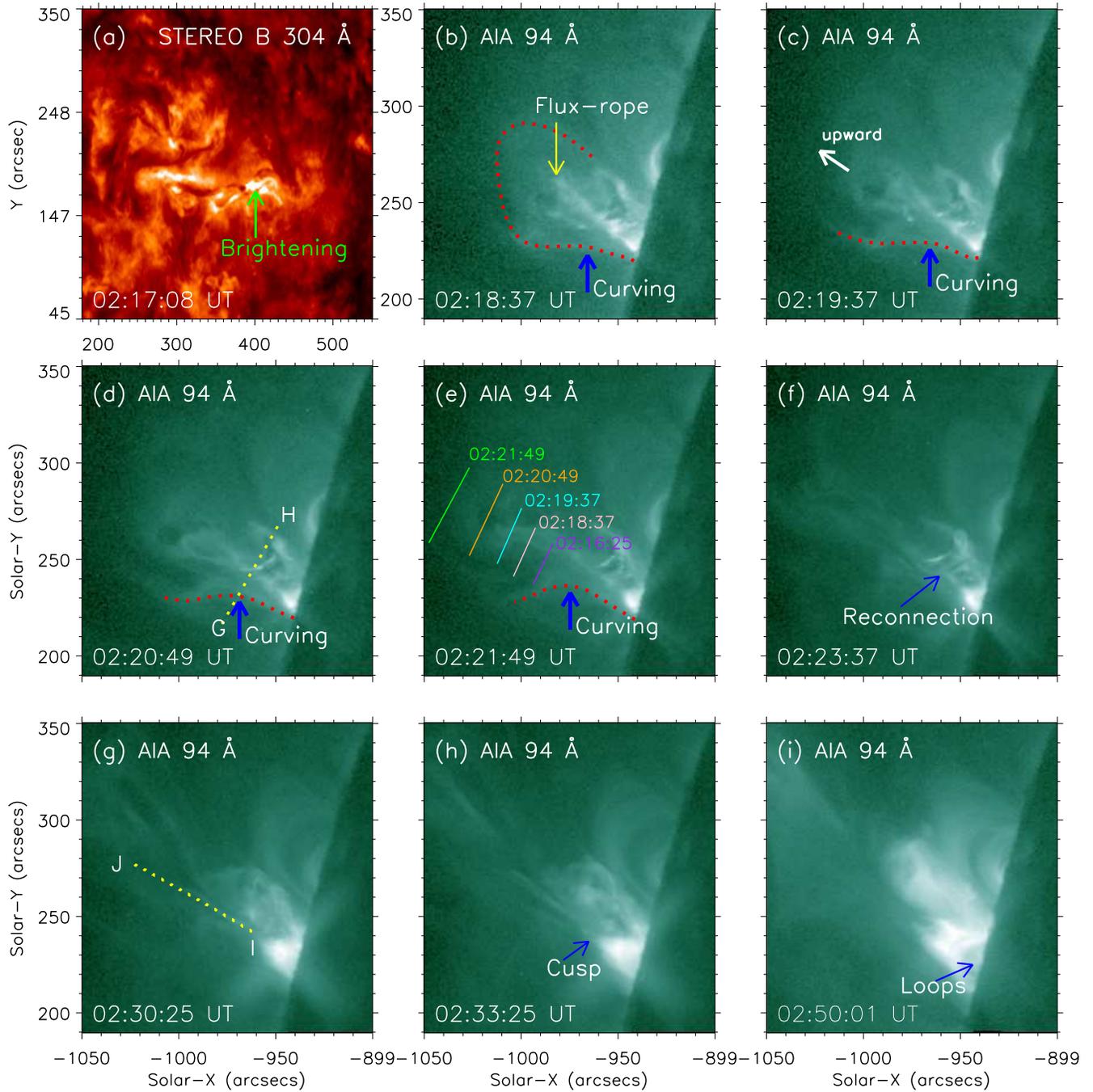}}
\caption{Reconnection process as observed in STEREO EUVI B 304 {\AA} (Panel a) and AIA 94 {\AA} (Panels b--i). The green arrow in Panel (a) points to the brightening around the footpoint. The red dotted lines in Panel (b) show the overlying large scale field lines while the yellow arrow points to the flux-rope structure. The thick blue arrows in Panels (b)--(e) mark the direction of destabilization of the field lines. In Panel (e), solid lines of different colors are drawn, in tangent to the upper edge of the flux-rope, at different times labelled correspondingly. Yellow dotted lines in Panels (d) and (g) indicate the two slices that are to be used to derive the inward and upward flow profiles. The thin blue arrows in Panels (f), (h) and (i) point to the reconnection position, cusp structure and the post-flare loops respectively. See also supplementary Movie 5.}\label{fig:xu5}
\end{figure*}

\begin{figure*}
{\includegraphics[width=1.0\textwidth]{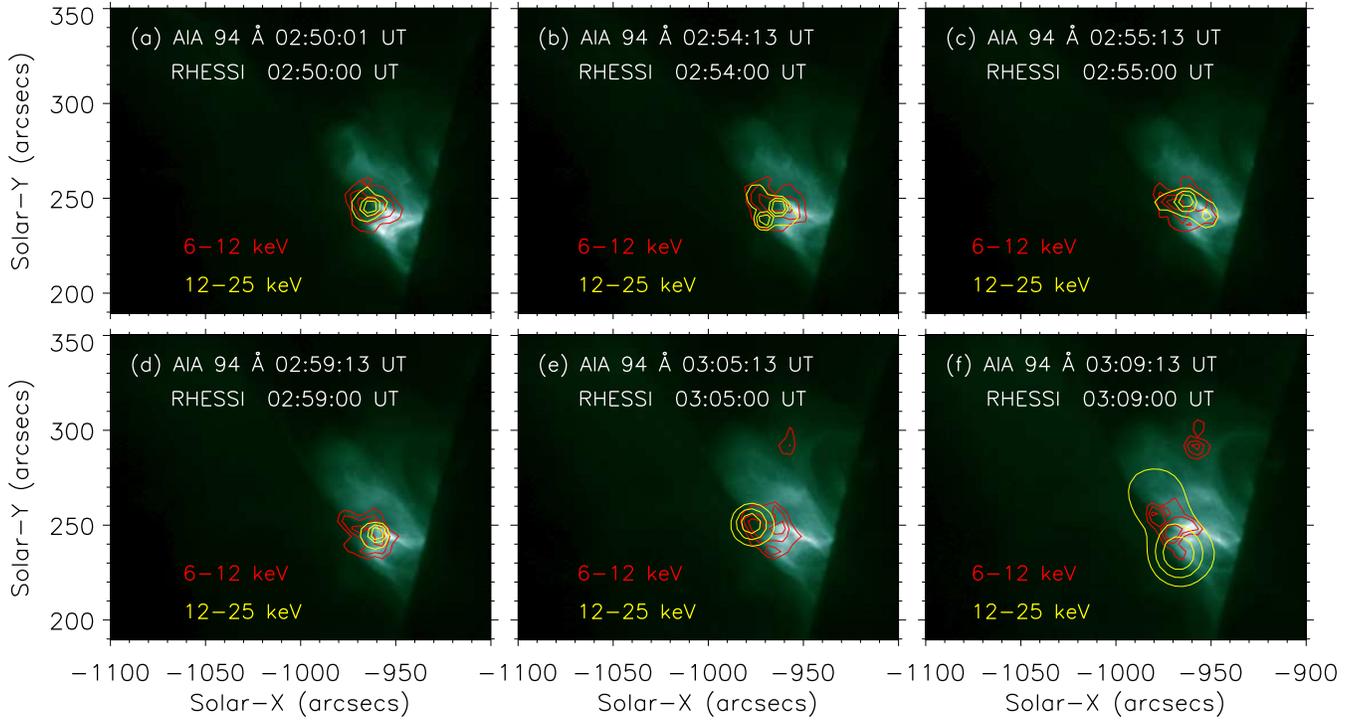}}
\caption{The AIA 94 \AA\ images show a nice cusp shape. Red and Yellow contours (40, 60 and 80\% of the maximum) outline an X-ray source in 6--12 keV and 12--25 keV, respectively, synthesized from RHESSI observations.}\label{fig:xu8}
\end{figure*}

\begin{figure*}
{\includegraphics[width=1.0\textwidth]{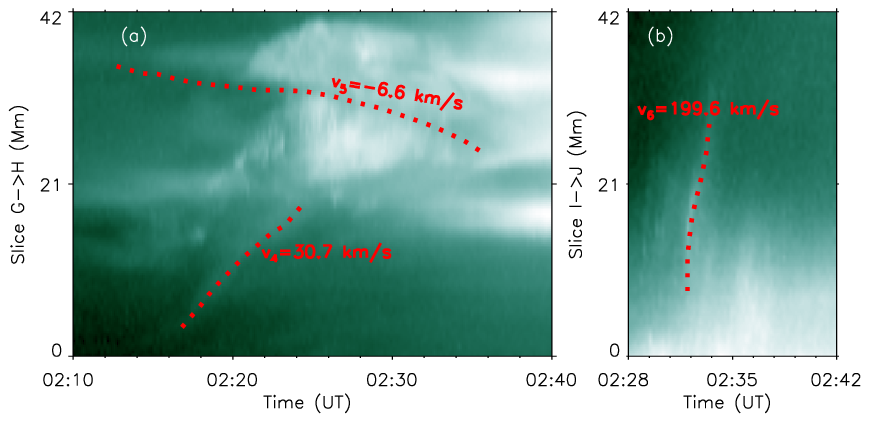}}
\caption{Time--distance plots, showing the variation of the intensity with the time along the slide from G$\rightarrow$H in Figure \ref{fig:xu5}(d) and I$\rightarrow$J in Figure \ref{fig:xu5}(g). The red dotted lines in Panel (a) track the inflows, with measured speeds of $\sim$ 30.7 km s$^{-1}$ and $\sim$ 6.6 km s$^{-1}$ respectively as labelled. The red dotted lines in Panel (b) denotes the upward flow and the measured speed  is $\sim$ 199.6 km s$^{-1}$). The Y-axis is the relative distance in the unit of Mm.}\label{fig:xu6}
\end{figure*}

We then investigated the kinematics of different structures within the prominence. We analyzed three slices as outlined by the cyan dotted lines in Figure \ref{fig:xu2}. The slice A$\rightarrow$B is of 485 pixels in length and 6 pixels in width, and our analysis start from 02:10:07 UT. We averaged the pixel intensity across the width and synthesized a time-distance map as plotted in Figure \ref{fig:xuth}(a). This reveals that there are three phases in the evolution: acceleration, constant speed, and deceleration. After 02:25:43 UT, a small portion of the prominence material kept rising up while the rest started to drain downward, which caused three typical motion tracks during the deceleration phase.

In order to calculate the speed and acceleration/deceleration, we took several data points along the green dotted lines and fit them with either a Polynomial fitting function $h=a+bt+ct^2$ (for the acceleration and deceleration phases) or a linear fitting function $h=a+bt$ (for the constant speed phase). From the best-fit functions, we derived the speed of the upflow increased to $\sim$65 km s$^{-1}$  at 02:19:55 UT and the acceleration was about 108 m s$^{-2}$ (a$_{1}$) during the acceleration phase 02:11:31--02:19:55 UT (5 points were used for fitting). The speed is $\sim $228 km s$^{-1}$ (v$_{1}$) during the constant speed phase  (15 points were used for fitting, from 02:20:55--02:25:43 UT).  It is evident that there is a sudden rapid acceleration during 02:19:55--02:20:55 UT. This time interval is short and by assuming a constant acceleration over this period, we deduced the acceleration to be~2715 m s$^{-2}$. This may mark the onset of the impulsive phase of the flare.

During the deceleration phase, the prominence material separated into several bunches. The speed of one fraction of the erupting prominence material (between two blue stars marked in Figure \ref{fig:xuth}(a)) decreased to $-$73 km s$^{-1}$ with a constant deceleration $\sim$391 m s$^{-2}$ (a$_{4}$) during 02:25:43--02:33:55 UT (10 points were used for fitting). However, after 02:33:55 UT, one portion of this plasma material still kept rising with a speed$\sim$20 km s$^{-1}$ (6 points were used for fitting), while most of prominence material had drained back to the Sun. Another fraction of the erupting prominence material decreased to $\sim$4.8 km s$^{-1}$ with a constant deceleration $\sim$217 m s$^{-2}$ (a$_{3}$) during 02:25:31--02:39:55 UT (10 points were used for fitting). At 02:43:43 UT, the loops erupted and the material fallen back to the surface of the Sun along the legs of prominence.

In order to determine the kinematics of the falling materials, we placed a slice E$\rightarrow$F along the prominence leg (cyan dotted lines in Figure \ref{fig:xu2}(e)). It is of 354 pixels in length and 10 pixels in width, and the space-time map has been synthesized  since 02:40:07 UT. Similar to the technique applied in the aforementioned kinematical calculations, we took the average value of intensity across the width to make the time-distance plot as presented in Figure \ref{fig:xuth}(c). We found several bright thread structures which indicate the uninterrupted motion of the plasma along the chosen slice. We chose one typical bright thread to estimate the velocity (green dotted line).  The plasmas fell with a constant speed $-$192 km s$^{-1}$(v$_{3}$) (15 points were used for fitting). We placed slice C$\rightarrow$D to study the kinematic of the upward flow in Figure \ref{fig:xu2}(d). It is of 158 pixels in length and 10 pixels in width, and the space-time map has been synthesized since 02:30:31 UT. We took the average value of intensity across the width to make time-distance plot in Figure \ref{fig:xuth}(b). The inferred velocity is $\sim$137 km s$^{-1}$(v$_{3}$) (10 points were used for fitting). This upward flow can also be seen in AIA hot channels 94 \AA\ and 131 {\AA}.

\subsection{The twist evolution}

The twist and writhe (see definitions in \citet{Tor10}) was clearly seen in all EUV channels recorded by AIA at around 02:20 UT. In order to make an in-depth investigation of the evolution of the twist, we extracted a rectangle region as denoted by the white box in Figure \ref{fig:xu2}(b). The twist of the rope gradually becomes evident from 02:18:35 UT in Figure \ref{fig:xu4}(a). As the prominence started rising and rotating, two crossing threads were apparent in AIA 171 \AA\ image at 02:20:35 UT, and the writhe was clearly seen in Figure \ref{fig:xu4}(b).  At nearly the same time, another writhed shape was evident in 131 \AA\ image, one of the very hot EUV passbands (Figure \ref{fig:xu4}(c)). A sketch of twist evolution is shown in Figure \ref{fig:xu11}(a) and (b). It is interesting that half of this structure appears to be very bright and narrow,  while the other half appears to be much wider and very faint and only presents in AIA 94 \AA\ and 131 {\AA} images (Figure \ref{fig:xu4}(d)). We inferred the prominence composed of both hot and cool plasma based on this phenomenon, although the reason of its formation is not very clear. This hot structure presents twisted or writhed axis in accordance with the general property of the flux rope investigated in \citet{Che11} and \citet{Zha12}. So we suggest that it is a flux rope.  After 02:22:47 UT, the clockwise rotation (green dotted lines) of the flux rope was observed with the two crossing threads separated gradually (Figure \ref{fig:xu4}(e) and (f)). The rotation direction is outlined by green arrows in the sketch (Figure \ref{fig:xu11}(c) and (d)). The whole evolution process can also be seen in supplementary Movie 4

From image at one time such as Figure \ref{fig:xu4}(b), we can see two bright threads crossing each other three times, which indicates that the twist of the prominence was at least 3$\pi$ (1.5 turns). To get an estimate of the twist number from the dynamics, we placed a slit perpendicular to the prominence (blue dashed line in Figure \ref{fig:xu4}(b)) and got its time-distance plot (Figure \ref{fig:xu12}). \textbf{A similar method was used by \citet{Ryu08}.} \textbf{The period $P$ is $\sim 20$ minutes which is the time between the maxima of the intensity outline curve (between two green asterisks marked in Figure \ref{fig:xu12}).}  \textbf{The speed $v_{a}$ ($13.2$ km/s) was estimated by linear fitting using the points between two cyan asterisks, and $v_{b}$ ($15.9$ km/s) was estimated using the points between green and blue asterisks. The} average speed \textbf{$v_{p}$} is $\sim 14.6$ km/s. The pitch of the screw thread can be estimated as $h_{p}=P \times v_{p}=17.5$ Mm. The total length ($L$) of the twisted loop is approximate 55 Mm, estimated at 02:20:35 UT when the twist is most clear. The twist can be estimated as $L/h_{p}\approx3$ turns (6$\pi$), which exceeded the threshold of kink instability. At each crossing (indicated by the white arrows in Panel b), the upper thread was left skewed relative to the lower thread, indicating the helicity of the prominence should be negative, according to method proposed by \citet{Cha00}. \citet{Chen14} proposed another method to determine the chirality of an erupting filament on the basis of the skewness of the conjugate filament drainage sites, i.e., the right-skewed (left-skewed) drainage corresponds to sinistral (dextral) chirality. We applied this method to Figure \ref{fig:xu3}(e), the skew of the drainage sites (F2 and B) is left, corresponding to dextral chirality (negative helicity). The results using these two methods are consistent with each other, showing the chirality of this prominence was dextral and followed the hemispheric preference studied by \citet{Ou17}.

\subsection{Magnetic reconnection}

The brightening around the filament footpoints, first appearing at 02:17:08 UT in STEREO B 304 \AA\ image (Figure \ref{fig:xu5}(a)), is interpreted as the result of a magnetic reconnection. Since the evidences of magnetic reconnection can be more clearly seen in hot channels such as AIA 94 \AA\ and 131 \AA\ channels, we took a time-series of AIA 94 \AA\ images (See also supplementary Movie 5) to study the reconnection process here. The red dotted line in Figure \ref{fig:xu5}(b) outlines the overall large scale field line and the yellow arrow denotes the inside flux-rope. The flux-rope rose up, most-likely due to kink instability, in the direction marked by the white arrow in \ref{fig:xu5}(c). The outside large-scale field continues to expand and curves inward with the rising of the flux-rope (Figure \ref{fig:xu5}(b)--(e)). We used different color solid lines to mark the upper edge of the flux-rope at different times in Figure \ref{fig:xu5}(e).  The rising speed increased gradually, evident by  comparing the distance between different color lines (i.e., the distance between cyan (02:19:37) and orange (02:20:49 UT) lines being larger than that between cyan and pink (02:18:37 UT) lines). This phenomenon is coherent with the kinematic character inferred from AIA 304 \AA\ images. At 02:21:49, the flux-rope erupted, which caused a decrease in internal pressure. This pressure imbalance between the interior and exterior caused the overlying large-scale field lines curving to the reconnection region (pointed by the blue arrow in Figure \ref{fig:xu5}(f)). At 02:30:25 UT, an upward reconnection jet was observed along the yellow dotted line in Figure \ref{fig:xu5}(g). From 02:33:25 UT, a cusp (denoted by blue arrow in Figure \ref{fig:xu5}(h)) structure gradually formed below the reconnection site.

The source region of the X-ray emission, deduced from the RHESSI observations, is found to be spatially linked to the bright emission region in the EUV images (Figure \ref{fig:xu8}), confirming the energy is released due to the magnetic reconnection. The RHESSI coronal source appeared in the 6--12 keV and 12--25 keV energy range and indicates the presence of plasma at temperature $> 6$ MK near the reconnection site. At 02:54 UT and 02:55 UT, the coronal source in 12--25 keV separated into two sources which may indicate the reconnection site changed during the decline of eruption.  At 03:05 UT, another coronal source appeared in 6--12 keV. This source is above the nearby loop which indicates that the reconnection probably occurred between the prominence's field lines and the nearby fields.

We analyzed a slice G$\rightarrow$H (yellow dotted line in Figure \ref{fig:xu5}(d)) to study the reconnection inflow profile. It is of 95 pixels in length and 10 pixels in width, and the space-time map has been synthesized since 02:10:13 UT. We took the average value of intensity across the width to make time-distance plot in Figure \ref{fig:xu6}(a). The two red dotted lines indicate the oppositely-directed inflows with the speeds  $\sim$ 30.7 km s$^{-1}$ ($v_{4}$) and $\sim$ 6.6 km s$^{-1}$ ($v_{5}$).  Slice I$\rightarrow$J (yellow dotted line in Figure \ref{fig:xu5}(g)) is of 95 pixels in length and 10 pixels in width, and the space-time map has been synthesized  since 02:28:13 UT. We took the average value of intensity across the width to make time-distance plot presented in Figure \ref{fig:xu6}(b). The red dotted line indicates the upward flow with speed $\sim$ 199.6 km s$^{-1}$ ($v_{6}$). It is worthy of noting that the upward flow was also observed in AIA 304 \AA\ images, but the speed was relatively lower than that deduced from 94 \AA\ images. A higher speed of hot plasma might have been resulted from the magnetic reconnection.

This magnetic reconnection process can be explained using the model developed in \cite{Shib96}, but the magnetic structure is more complex than that employed in the model.

\begin{figure*}
{\includegraphics[width=1.0\textwidth]{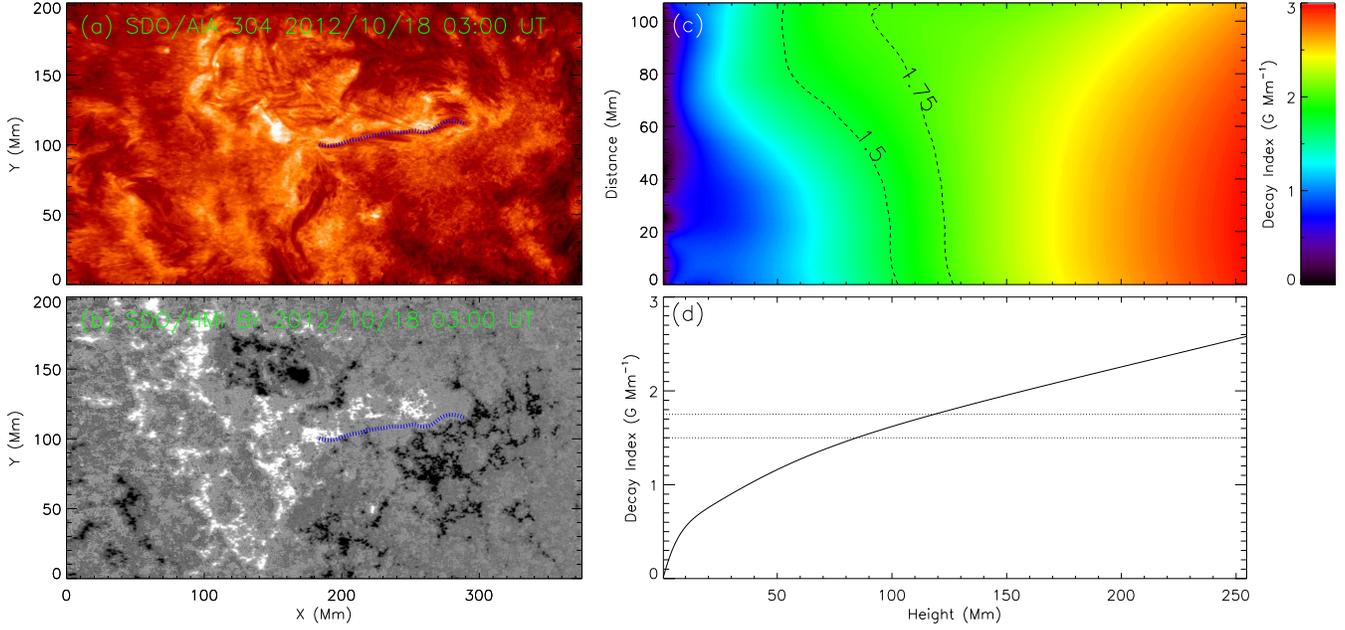}}
\caption{The newly formed filament at the same location as the investigated event after it turned to the front side of the Sun on 18 Oct. 2012. (a) SDO/AIA 304 \AA\ observation. (b) the vertical component of the magnetic field (with saturation limits of $\pm$500 G) observed by SDO/HMI. The target filament is outlined by the blue dotted curve in both panels. (c) the decay index above the filament with the x-axis and y-axis as the distance along the filament from its left end and the height above the photosphere, respectively. Two dashed curves are the contours of the decay index at levels of 1.5 and 1.75. (d) the average decay index over the filament, with the two horizontal dotted lines at levels of 1.5 and 1.75. }\label{fig:xu10}
\end{figure*}

\section{Conclusions and Discussions}

Through multi-wavelength diagnostics we observed the whole evolution process of a failed prominence eruption. The initial state is a stable filament on the solar disk as observed by EUVI/STEREO, while it appears as a prominence on solar limb in H$\alpha$ images. The filament/prominence then became unstable and depicted a rising motion. During this phase, the twist got converted to writhe. After the flare, the prominence rose in an accelerated manner, with reconnecting prominence threads and a cusp-shaped structure seen in hot channels of EUV images.

The twist of the filament was estimated approximate 6$\pi$ (3 turns), which exceeds the threshold of the kink instability. This suggests that this event may be triggered by the kink instability.

The largest rising  speed  was 228 km s$^{-1}$. There was a sudden rapid acceleration at the time of the flare onset, with an estimated value of 2715 m s$^{-2}$.

However, this eruption was not accompanied by a CME. The prominence material was seen to fall back to the chromosphere. A new filament was formed at almost the same location of the original one. The largest deceleration of one portion of the prominence was 391 m s$^{-2}$, whose value is even larger than the solar gravitational constant (g=274 m  s$^{-2}$ ).

During the eruption, the magnetic reconnection can be identified by the clearly seen inflows and nice cusp structure. The RHESSI X-ray emissions further confirmed the co-spatial X-ray emission, which signifies the energy release location of the reconnection.

The helical kink instability was regarded as an important triggering mechanism for solar eruptions \citep[e.g.,][]{Tor05, Rus05}. Whether the kink instability leads to a failed or eruptive event depends on the decrease of the overlying magnetic field with height \citep{Tor05}. Fast decreasing of the external magnetic field of a flux rope along height could result in the torus instability and finally result in the eruption of the flux rope, otherwise it will lead to a failed eruption. Define the decay index as $n=-\frac{d ln(B_{ex})}{d ln(z)}$ where $B_{ex}$ and $z$ are the external magnetic field strength and height respectively. Theoretical and observational studies have found the torus instability usually occur when the decay index is above a certain threshold between 1.5 to 1.75 \citep{Kli06, Fan10, Wan17}. Corresponding height above the solar surface where the decay index reaches the critical threshold is called the critical height.

In our case, the twist is clearly seen to be converting to writhe motion following the appearance of a kink instability, which confirmed that the helical kink instability can trigger the prominence/filament eruption. However, in the absence of magnetic field observation, it is not possible to infer whether or not torus instability took place during the eruption process. To determine the critical height of the filament eruption, we performed a potential field extrapolation using a Fourier Transformation method \citep{Alis81} from the vertical component of the vector magnetic field obtained from SDO/HMI active region patch (HARP) 2117 with coordinates transformed to a heliographic Cylindrical Equal-Area (CEA) projection \citep{Bob14} at 03:00 UT on 18 Oct. 2012 (Figure \ref{fig:xu10}(b)). Figure  \ref{fig:xu10}(a)) is the simultaneous SDO/AIA 304 \AA\ observation with the coordinates transformed to the same projection. The target filament is outlined by the blue dotted curve in both panels. The newly formed filament stayed relatively stable after the investigated event after it turned to the front side of the Sun. Following the solar rotation, the filament's center was approximately 40$^\circ$ east of the central meridian at this time, allowing more accurate observations of the photospheric magnetic field than earlier times. The decay index above the filament was then calculated from the potential field extrapolation, showing in Figure \ref{fig:xu10}(c) with the x-axis and y-axis as the distance along the filament from its left end and the height above the photosphere, respectively. Two dashed curves are the contours of the decay index at levels of 1.5 and 1.75. Figure \ref{fig:xu10}(d) shows the average decay index over the filament, with the two horizontal dotted lines at levels of 1.5 and 1.75. \textbf{The} average decay index reaches the critical points of 1.5 and 1.75 at heights of $\sim$85 Mm and $\sim$118 Mm, respectively. A portion of material of investigated event started to fall at height $\sim$100 Mm which reached the lower critical height for torus instability. But the prominence eruption didn't lead to CME. \textbf{There is a 4-day} gap between the investigated event and the calculation of decay index, and the magnetic field may change during this period. \citet{Hab14} pointed out that the helical  wavy  patterns appear as a natural byproduct of the inherent dynamics of prominences that do not necessarily lead to CMEs, even when prominences erupt. The restraining force of the overlying flux and the cancelation of the upward Lorentz force were also suggested as one mechanism for the failed eruption \citep[e.g.,][]{Xue16b}. Since we obtained the maximum deceleration speed to be 391 m s$^{-2}$, which is larger than the solar gravitational constant, it may further suggest that this prominence eruption failed due to the inward magnetic tension force.

The magnetic reconnection process was first seen as the brightening observed around the footpoints in EUVI/STEREO images, which first triggered the flare eruption and then accelerated the filament. During the rising period, the magnetic reconnection between the prominence threads and overlying field occurred, evident by AIA 131 \AA\ and 94 \AA{} observations. The reconnection rate is an important parameter for magnetic reconnection, defined as the reconnected magnetic flux per unit time. The dimensionless form of the reconnection rate is the Alfv\'{e}n Mach number $M_{A}=V_{in}/V_{A}$, where $V_{in}$ is the inflow speed of reconnection and $V_{A}$ is the Alfv\'{e}n speed. For the studied case, the inflow speed is 6.6--30.7 km s$^{-1}$ and the outflow speed is 199.6 km s$^{-1}$. Assuming the outflow speed to be equal to the Alfv\'{e}n speed,  we can estimate the reconnection rate as $M_{A} \approx V_{in}/V_{out}$ $\approx$ 0.03--0.15. Our estimation is roughly consistent with \citet{Pet64}'s model ($M_{A}$=0.01--0.1) and the estimation for several observed solar events, e.g., 0.01--0.23 \citep{Lin05}, 0.055--0.2 \citep{Tak12}, 0.05--0.5 \citep{Su13} and 0.08--0.6 \citep{Xue16b}.

 In recent years, a series of comparative studies about the eruptive, confined (failed) and partial eruption have been done \citep[e.g.,][]{Shen11, zhqm15, Liu18}. It is \textbf{shown} that some parameters, i.e., the decay index, field strength at low corona etc., had no significant difference for failed and successful eruption \citep{Shen11}. \cite{llj18} suggested that failed eruption may be the result of a combination of several mechanisms including the weaker non-potentiality in the core region, smaller Lorentz force impulse during the eruption, and the local torus-stable region in the coronal magnetic fields. The trigger mechanism and energy release process of solar eruption is still not very clear and requires further study.

\acknowledgments We acknowledge the use of data of SMAT/HSOS, AIA/SDO, EUVI/STEREO and RHESSI. This work is supported by the National Natural
Science Foundation of China (Grant Nos. 11703042, 11911530089, U1831107, 11673033, 11427901, 11573012, U1731241, U1531247, 11773038, 11973056), Strategic Priority Research Program on Space Science of the Chinese Academy of Sciences (Grant Nos. XDA15052200, XDA15320302) and the Provincial Natural Science Foundation of Shandong (Grant No. ZR2018MA031).

\end{document}